\documentclass[onecolumn,amsmath,amssymb,aps]{revtex4-2}
\usepackage{physics}
\usepackage{graphicx}
\usepackage{dcolumn}
\usepackage{bm}
\usepackage{caption}
\usepackage{bbold}
\usepackage{subcaption}
\usepackage{xcolor}
\usepackage{tikz}
\usetikzlibrary{decorations.pathmorphing}
\tikzset{zigzag/.style={decorate, decoration=zigzag}}

\definecolor{darkgreen}{HTML}{006622}
\newcommand{\comment}[1]{}

\begin{document}

\title{Measurement-induced nonlocality for observers near a black hole}


\author{Adam Z. Kaczmarek$^{1}$}
\author{Dominik Szcz{\c{e}}{\'s}niak$^{1}$}\email{d.szczesniak@ujd.edu.pl}
\author{Sabre Kais$^{2}$}


\affiliation{$^1$Department of Theoretical Physics, Faculty of Science and Technology, Jan D{\l}ugosz University in Cz{\c{e}}stochowa, 13/15 Armii Krajowej Ave., 42200 Cz{\c{e}}stochowa, Poland}
\affiliation{$^2$Department of Chemistry, Department of Physics and Astronomy, and Purdue Quantum Science and Engineering Institute, Purdue University, 47907 West Lafayette, Indiana, United States of America}


\date{\today}


\begin{abstract}
We present a systematic and complementary study of quantum correlations near a black hole by considering the measurement-induced nonlocality (MIN). The quantum measure of interest is discussed on the same footing for the fermionic, bosonic and mixed fermion-boson modes in relation to the Hawking radiation. The obtained results show that in the infinite Hawking temperature limit, the physically accessible correlations does not vanish only in the fermionic case. However, the higher frequency modes can sustain correlations for the finite Hawking temperature, with mixed system being more sensitive towards increase of the fermionic frequencies than the bosonic ones. Since the MIN for the latter modes quickly diminishes, the increased frequency may be a way to maintain nonlocal correlations for the scenarios at the finite Hawking temperature.
\end{abstract}


\maketitle


\section{Introduction}


The behavior of quantum information in the relativistic settings gained significant attention over the recent years \cite{peres2002, ahn2003, terashima2003, hu2018, lanzagorta2014} and the advent of this interest can be associated with the pioneering work of Peres, Scudo and Terno, showing that the spin entropy is not Lorentz invariant \cite{peres2002}. Since then, the quantum information theory faced important revisions in contact with the relativistic world. This led to the creation of the relativistic quantum information (RQI) domain and emergence of the new effects \cite{fuentes2013}. For example, it was shown that the entanglement phenomenon \cite{amico2008, horodecki2009} is a observer-dependent entity \cite{fuentes2005, alsing2006} and fidelity of teleportation is affected by the uniformly accelerated observers \cite{alsing2004a}. This is to say, both entanglement and teleportation appears to be sensitive to the so-called Unruh effect, saying that the notion of vacuum (as measured by an observer) depends on the observer's space-time path \cite{unruh1976, hollands2015}. As a result, the RQI arise as an interdisciplinary field that may cover various aspects of relationship between gravity and the quantum world. This includes, but is not limited to, the security in quantum cryptography against {\it gravitational attacks} \cite{plaga2006}, efficient simulations of quantum systems when relativistic effects are required \cite{veis2012}, fast and precise quantum information processing at large distances \cite{sidhu2021}, quantum communication and metrology in gravitational fields \cite{ahmadi2014, kohlrus2017} as well as the fundamental considerations {\it e.g.} investigations of the gravitationally-induced entanglement \cite{krisnanda2021}.

In the RQI field there is also a particular interest in the quantum correlations and their quantifiers, with a special attention given to the nonlocality \cite{friis2011, hu2018}. The rationale behind this is based on the nonlocality being some form of a correlation. As a physical concept, the nonlocality goes beyond Bell's theorem and there is possibility to obtain {\it nonlocality without quantumness} and {\it nonlocality without entanglement} \cite{walgate2002,bennet1996,luo2011}. Thus, related investigations are conducted not only for the conventional Bell-like non-locality measures \cite{genovese2005} but also for more general concepts initiated recently \cite{hu2018}. One of such novel measures is known as the measurement-induced nonlocality (MIN) and amounts for the {\it maximum global effect caused by locally invariant measurements} \cite{luo2011}, being somewhat dual to the quantum discord \cite{ollivier2001}. Specifically, the MIN was originally introduced via the Hilbert-Shmidt norm as the quantification of nonlocality from a geometric perspective, for consideration of the quantum information processing problems \cite{luo2011}. Later on, this approach was developed further based on the relative entropy \cite{xi2012}, the trace norm \cite{Hu2015}, the fidelity measure \cite{muthagenesan2017} or the skew information \cite{zhang2018} in order to avoid some of the limitations of its initial formulation {\it e.g.} the noncontractivity. In a result, the MIN may currently serve as a suitable platform for the analysis of bi- and multi-partite systems, allowing comparison of various correlations as well as the discussion of related quantum effects such as the teleportation \cite{bennet1993} or the quantum steering \cite{brunner2010}.

In the relativistic frame, the MIN was so far studied for the Unruh observers \cite{tian2013} and in the dilaton space-time setup \cite{he2016}. It was shown that at the infinite Unruh temperature limit, the correlations are non-vanishing for fermions, contrary to the bosonic modes \cite{tian2013,he2016}. Although mentioned studies considered only limited cases, they pointed to the importance of addressing nonlocality and mutual information in the non-inertial systems, similarly to the investigations based on other than the MIN measures \cite{friis2011,smith2011,doukas2013}. They have also shown that MIN is a general framework that may capture aspects of nonlocality not accessible by means of the Bell-inequality \cite{he2016}. In the light of these results, an essential objects to investigate MIN for the non-inertial observers appears to be a black holes with their quantum-gravitational foundations \cite{navarro2005,lanzagorta2014,kiefer2012}. The pivotal role in this respect is played by the Hawking radiation, a manifestation of the quantum mechanics in the space-time of a black hole (BH) \cite{hawking1974}, which is somewhat related to the mentioned Unruh effect \cite{davies1975,fulling1973,unruh1976}. By employing this phenomenon it was already possible to show that quantum correlations and similar effects are core for the BH thermodynamics and the information loss problem \cite{hawking1975,terashima2000}. Moreover, the Hawking radiation allowed studies on the behavior of quantum information in the vicinity of a BH such as the investigations of entanglement and teleportation in the background of the static \cite{pan2008} or the high-dimensional and rotating BHs \cite{ge2008}. In what follows, the BHs established themselves as a source of strong gravity field that can be employed for various investigations within the RQI field, {\it e.g} toward future advances in quantum computation and information processing \cite{friis2013,blasco2015}.

In this work, we attempt to present our contribution to the above domain of research by providing the general, unified and comprehensive description of the MIN in a BH space-time. Therefore, the results presented here are relevant to a large class of BHs, including the nonsingular ones \cite{hayward2006}. For this purpose, we conventionally consider setup composed of the Alice and Bob archetypes, where the first observer is away from the BH and the latter one within its space-time. They both share maximally entangled Werner state. Such setup, allows us to extend previous discussions given in \cite{tian2013,he2016}, by describing the MIN not only for the {\it homogenous} fermion-fermion or boson-boson correlations but also in the mixed fermion-boson case, on the same footing. Moreover, the analysis is conducted here directly with respect to the Hawking radiation by considering the MIN dependence on the Hawking temperature ($T_H$) for the Werner states of choice. To this end, the generality of our approach additionally permits us to investigate MIN as a function of the fermionic/bosonic frequency, in analogy to the similar studies conducted previously for the entanglement \cite{ge2008}.

The work is organized as follows: in Sec. II we describe vacuum structure of for the fermionic and bosonic fields in the BH space-time. Next, in Sec. III we derive MIN for the three correlation scenarios of interest. Finally, we conclude our discussion with some pertinent remarks and perspectives for future research.


\section{Vacuum structure of a black hole}

The starting point for the analysis is the static and spherical metric of the following form \cite{hobson2006,das2011}:
\begin{align}
    ds^2=-A(r)dt^2+B(r)^{-1}dr^2+r^2(d\theta^2+\sin^2d\varphi^2),
    \label{eq1}
\end{align}
with $A(r)=B(r)=f(r)$, where $f(r)$ is the radial function of choice, whose form varies depending on the considered BH type. To keep our considerations as general as possible, we do not assume any specific form of the $f(r)$ function and discuss the MIN in the {\it model-independent} manner. 

Afterwards, the line element can be rewritten in terms of the Edington-Finkelstein coordinates \cite{penrose1965}:
\begin{align}
    u=t-\int \frac{1}{f(r)}\text{d}r,\;\;\;\; v=t+\int \frac{1}{f(r)}\text{d}r,
     \label{eq2}
\end{align}
to permit propagation of signals through the space-time \cite{navarro2005}.

We also note that for any asymptotically flat BH space-time, the event horizon will be the Killing horizon, which is a null hypersurface whose vector field (the Killing vector field) is null at the surface \cite{faraoni2015}. In what follows, the normal vector ($k^\mu$) will be the Killing vector that satisfies \cite{faraoni2015}:
\begin{align}
    \mathcal{L}_k g_{\alpha\beta}=\nabla_\beta k_\alpha +  \nabla_\alpha k_\beta,
     \label{eq3}
\end{align}
where $\nabla_\beta$ denotes the usual covariant derivative, while $\mathcal{L}_k$ is the Lie derivative along the $k^\mu$. The surface gravity for the Killing vector relates the covariant directional derivative of the horizon's vector (along itself) to $k^\mu$ {\it i.e.} $k^\mu \nabla_ \mu k^\nu=\kappa k^\nu$ \cite{damour2004}. Note that $\kappa$ stands for the surface gravity of the horizon and in the Planck's unit system ($c=\hbar=G=k_B=1$) its unit is $1/t_P^2$, where $t_P^2$ is the Planck time.
With this in mind, when a BH radiates, the Hawking temperature will be given by \cite{navarro2005}:
\begin{align}
    T_H=\frac{\kappa}{2\pi}.
     \label{eq4}
\end{align}
We remind that Eq. (\ref{eq4}) is pivotal to our analysis, since it relates general relativity to the quantum field theory and allows inclusion of the BH characteristics in our theoretical framework.




\subsection{Fermionic modes}


In order to describe vaccum state of the cruved spacetime for fermions, one can start with the following Dirac equation:
\begin{align}
    (i\gamma^a e^\mu_aD_\mu-m)\psi=0,
     \label{eq5}
\end{align}
with $D_\mu=\partial_\mu-\frac{i}{4}\omega^{ab}_\mu \sigma_{a b}$, $\sigma_{ab}=\frac{i}{2}\{\gamma_a,\gamma_b\}$, $e^\mu_a$ being {\it vierbein} and $\omega^{ab}_\mu$ the spin connection. The solutions of such Dirac equation in regions $I$ (the Universe) and $II$ (a black hole) ({\it i.e.} outside and inside the event horizon ($r=r_H$), respectively) are:
\begin{align}
    \psi_\textbf{k}^{I+}=\vartheta e^{-i\omega_i u}, \;\;\; \psi_\textbf{k}^{II+}=\vartheta e^{i\omega_i u},
     \label{eq6}
\end{align}
where $\omega_i$ stands for the monochromatic frequency of the Dirac field (having the unit $1/l_P$, where $l_P$ is the Planck length) and $\vartheta$ is the 4-component Dirac spinor composed of the spinorial spherical harmonics.


Since solutions of Eq. (\ref{eq6}) are analytic inside and outside the event horizon, they completely span the orthogonal basis. Thus, the Dirac field can be written as \cite{alsing2004b}:
\begin{align}
    \Psi_{out}=\sum_i \int d\textbf{k} \Big[a^I_\textbf{k}\psi^{I+}_\textbf{k}+    a^{II}_\textbf{k}\psi^{II+}_\textbf{k}+ b_\textbf{k}^{\dagger I}\psi^{I-}_\textbf{k}+b_\textbf{k}^{\dagger II}\psi^{II-}_{\textbf{k}}\Big],
 \label{eq7}
\end{align}
where $a^I_\textbf{k}$ ($a^{II}_\textbf{k}$) and $b^{\dagger I}_\textbf{k}$ ($b^{\dagger II}_\textbf{k}$) are the fermion annihilation and antifermion creation operators for the exterior (interior) region of a BH. Note that summation in Eq. (\ref{eq7}) is over the frequencies (see Eq. (\ref{eq6})).

In addition to the above, it is required to describe quantum fields near the event horizon. For this purpose, after discussing the vacuum structure of a BH space-time \cite{damour1976,navarro2005}, we follow the formalism developed in \cite{alsing2004b, alsing2006}. Specifically, by using the light-like Kruskal coordinates \cite{kruskal1960,hobson2006,hemming2001}:
\begin{align}\nonumber
    &U=-\frac{1}{\kappa}e^{-\kappa u},\;\;\; V=\frac{1}{\kappa}e^{\kappa v},\;\; (r>r_H), \\
    &U=\frac{1}{\kappa}e^{\kappa u},\;\;\; V=-\frac{1}{\kappa}e^{-\kappa v},\;\; (r<r_H), 
     \label{eq8}
\end{align}
we rewrite the solutions of the Dirac equation as follows:
\begin{align}
    \psi_\textbf{k}^{I+}=\vartheta( U\kappa) e^{-i\omega_i/\kappa}, \;\;\; \psi_\textbf{k}^{II+}=\vartheta (-U\kappa)e^{i\omega_i /\kappa}.
     \label{eq9}
\end{align}
Next, to construct complete basis for the positive energy modes, the analytic continuation is done for Eq. (\ref{eq9}) by using the Damour's technique \cite{damour1976,ge2008}. We arrive at the following solutions:
\begin{align}
    &\psi_\textbf{k}^{I+}=\vartheta(\kappa U) e^{-i\omega_i/\kappa}=\Theta(-U)(-U\kappa)^{i\omega_i/\kappa},\;\;\;\;\psi_\textbf{k}^{II+}=\vartheta (-U\kappa)e^{i\omega_i /\kappa} =\Theta(U)(U\kappa)^{-i\omega_i/\kappa},
     \label{eq10}
\end{align}
where $\Theta(\pm U)$ is the Heavside step function. After that, the complete basis of interest can be expressed as the following linear combination of the modes:
\begin{align}
    \chi^1_{k}&=\psi_\textbf{k}^{I+}+ \widetilde{ \psi}_\textbf{k}^{II-}=\Theta(-U)(-U \kappa) ^{i\omega_i/\kappa}+e^{\pi \omega_i / \kappa}\Theta(U)(U\kappa)^{i\omega_i/\kappa},
    \label{eq11}
    \\ 
    \chi^2_{k}&=\psi_\textbf{k}^{I-}+ \widetilde{ \psi}_\textbf{k}^{II+}=\Theta(-U)(-U\kappa)^{-i\omega_i/\kappa}+e^{-\pi \omega/\kappa}\Theta(U)(U\kappa)^{-i\omega_i /\kappa},
    \label{eq12}
\end{align}
with:
\begin{align}
    \widetilde{ \psi}_\textbf{k}^{II+}=e^{-\pi\omega_i}(U\kappa)^{-i\omega_i/\kappa}.
     \label{eq13}
\end{align}
Hence, the complete basis for the normalized modes reads:
\begin{align}\nonumber
    \zeta^{I+}_\textbf{k}&=e^{\pi \omega_i/( 2\kappa)} \chi^1_{k} =e^{\pi \omega_i/( 2\kappa)}\Theta(-U)(-U \kappa) ^{i\omega_i/\kappa} +e^{-\pi \omega_i /(2 \kappa)}\Theta(U)(U\kappa)^{i\omega_i\kappa}\\ &=e^{\pi \omega_i/( 2\kappa)}\psi_\textbf{k}^{I+}+e^{-\pi \omega_i /(2 \kappa)}{\psi}_\textbf{k}^{II-},
    \label{eq14}
    \\ \nonumber \zeta^{II+}_\textbf{k}&=e^{-\pi \omega_i/( 2\kappa)}\chi^2_{k}= e^{-\pi \omega_i/( 2\kappa)}\Theta(-U)(-U\kappa)^{-i\omega_i/\kappa}+ e^{\pi \omega/(2\kappa)}\Theta(U)(U\kappa)^{-i\omega_i/\kappa}\\    &=e^{-\pi \omega_i/( 2\kappa)}{\psi}_\textbf{k}^{I-}+e^{\pi \omega/(2\kappa)}{\psi}_\textbf{k}^{II+}.
    \label{eq15}
\end{align}
Finally, based on the Eqs. (\ref{eq14})-(\ref{eq15}), the outgoing Dirac fields can be expanded in the Kruskal space-time:
\begin{align}
    \Psi_{out}&=\sum_i d\textbf{k}\frac{1}{\sqrt{2\cosh(\pi \omega_i/\kappa)}}\Big[ c^I_\textbf{k}\zeta^{I+}_\textbf{k}+c^{II}_\textbf{k}\zeta^{II+}_\textbf{k} +d^{I\dagger}_\textbf{k}\zeta^{I_-}_\textbf{k}+d^{II\dagger}_\textbf{k}\zeta^{II_-}_\textbf{k}\Big],
    \label{eq16}
\end{align}
where $c_\textbf{k}$'s and $d_\textbf{k}^\dagger$'s are creation and annihilation operators applied to the Kruskal vacuum.

To this end, via the Bogoliubov transformation, the creation and annihilation operators in the BH and the Kruskal space-times can be related to each other \cite{navarro2005}:
\begin{align}
    c^I_\textbf{k}=\alpha a^I_\textbf{k} -\beta b^{II\dagger}_\textbf{k},
\label{eq17}
\end{align}
with the following Bogoliubov coefficients:
\begin{align}
    \alpha=\frac{1}{(e^{-\omega_i /T_H}+1)^{1/2}},\;\;\; \beta=\frac{1}{(e^{\omega_i /T_H}+1)^{1/2}},
     \label{eq18}
\end{align}
Therefore, the vacuum and excited states of the Minkowski space-time will be related to the Kruskal's one via the following relationships:
\begin{align}    \ket{0}^+_{\textbf{k}}=\alpha\ket{0_\textbf{k}}_I^+\ket{0_{-\textbf{k}}}^-_{II}+\beta \ket{1_\textbf{k}}_I^+\ket{1_{-\textbf{k}}}^-_{II},\;\;\;\;
    \ket{1_{\textbf{k}}}^+=\ket{1}^+_{I}\ket{0_{-\textbf{k}}}_{II}^-.
    \label{eq19}
\end{align}
%


\subsection{Bosonic modes}


In case of the bosonic field one is required to deal with the massless scalar field in the curved space-time, being the solution of the Klein-Gordon equation \cite{navarro2005,pan2008}:
\begin{align}
    \frac{1}{\sqrt{-g}} \frac{\partial}{\partial x^\mu}\Big(   \sqrt{-g} g^{\mu\nu} \frac{\partial \phi}{\partial x^\nu}\Big)=0.
     \label{eq20}
\end{align}
In details, by considering Eq. (\ref{eq20}) near the event horizon, we can arrive with the incoming wave function, which is analytic for the entire space-time and given as:
\begin{align}
     \phi^{in}=Y_{lm} e^{i\Omega u},
     \label{eq21}
\end{align}
Similarly, the outgoing wave functions for the regions $I$ and $II$ of the event horizon are \cite{pan2008}:
%
\begin{align}
    \phi^I_\Omega=Y_{lm}e^{i\Omega u}, \;\;\;\; \phi^{II}_\Omega=Y_{lm}e^{-i\Omega u}.
     \label{eq22}
\end{align}
In Eqs. (\ref{eq21})-(\ref{eq22}), $Y_{lm}(\theta,\varphi)$ denotes scalar spherical harmonic and $\Omega$ is the bosonic frequency of the mode with the units $1/l_P$.
Analogously to the fermionic case, the $\phi^I$ and $\phi^{II}$ solutions span orthogonal basis for the Klein-Gordon field \cite{pan2008}:
\begin{align}
\Phi_{out}=\sum_{lm} \int d\Omega
 \Big[a^I_\Omega\phi^{I}_\Omega+    a^{II}_\Omega\phi^{II}_\Omega+ b_\Omega^{\dagger I}\phi^{I*}_\Omega+b_\Omega^{\dagger II}\phi^{II*}_{\Omega}\Big],
 \label{eq23}
\end{align}
where $a^I_\omega$ ($a^{II}_\Omega$) and $b^{\dagger I}_\Omega$ ($b^{\dagger II}_\Omega$) are the bosonic annihilation and creation operators for the exterior (interior) region of the considered BH. Note again, that summation in Eq. (\ref{eq23}) is over the spherical harmonics (see Eq. (\ref{eq22})).

By introducing the following abbreviations:
\begin{align}
    \cosh{r}=\frac{1}{\sqrt{1-e^{-\Omega/T_H}}},\;\;\;\; \tanh{r}=e^{-\Omega/T_H},
     \label{eq24}
\end{align}
and
\begin{align}
    \tanh{r} = t,\;\;\;\;\;\cosh{r}= 1/(\sqrt{1-t^2}),
 \label{eq25}
\end{align}
one can write down the vacuum and excited bosonic modes in a more convenient form:
\begin{align}
    \ket{0}_\textbf{k}&=\sqrt{1-t^2}\sum_{n=0}^\infty t^n \ket{n}_I \ket{n}_{II},\;\;\;\; \ket{1}_{\textbf{k}}=(1-t^2)\sum_{n=0}^\infty \sqrt{n+1}t^n\ket{n+1}_I \ket{n}_{II}.
\label{eq26}
\end{align}
%


\section{Measurement-induced nonlocality}


The main setup employed here is conventional and follows scenarios originally presented in \cite{fuentes2005, ge2008, he2016}. In details, we consider two observers: first Alice equipped with the particle detector sensitive only to mode $\ket{n}_A$ and second Bob detecting only mode $\ket{n}_B$ by using his device. In such a framework, both observers initially share an entangled state for the same event in the flat region of Minkowski spacetime. Next, it is assumed that Alice remains stationary at the asymptotically flat region of the space-time ($r \rightarrow \infty$), while Bob first freely falls in the direction of a BH and at some point starts to hover near the event horizon. As a result, Bob becomes affected by the thermal {\it{bath}} of the particles associated with the Hawking radiation \cite{hawking1974,davies1975,unruh1976}. Therefore, to describe what Bob will detect, mode $\ket{n}_B$ is specified in the coordinates of a BH. 

Here it is instructive to elucidate several aspects of the above setup. First, since entanglement is sensitive to the environment, we note that Bob's trajectory may alter description of the state. This potential issue is avoided by assuming that Bob is free-falling toward a BH, slowly decelerating before becoming stationary \cite{ge2008}. Second, it is noted that the states are constructed here in analogy to the so-called single-mode approximation \cite{alsing2003}. Although such approach has its limitations, it has been found to recover most important qualitative features of more sophisticated models \cite{bruschi2010}. To this end, we remark that additional details of the method employed here can be found in previous studies such as \cite{he2016, pan2008} or in the recent review by Hu et al. \cite{hu2018}.



\subsection{Fermionic modes}


We begin with the Werner state as our initial state \cite{werner1989,he2016}:
\begin{align}
    \rho_{AB}= \rho_{A B_I B_{II}}=\eta\ket{\phi^+}\bra{\phi^+}+\frac{1-\eta}{4} \mathcal{I}, 
    \label{eq27}
\end{align}
which is based on the following Bell state:
\begin{align}
    \ket{\phi^+}&=\frac{1}{\sqrt{2}}\big(\ket{00}+\ket{11}\big)=\frac{1}{\sqrt{2}}\big(\alpha \ket{0_A0_I0_{II}}+\beta \ket{0_A 1_I 1_{II}}+\ket{1_A 1_I 0_{II}}\big),
    \label{eq28}
\end{align}
and $\mathcal{I}=\ket{00}\bra{00}+\ket{11}\bra{11}$. We remark that $\rho_{AB}$ is in the Minkowski basis $\ket{00}, \ket{01}, \ket{10}, \ket{11} $ ($\ket{ab}=\ket{a}_A\ket{b}_B$). Letters $A$ and $B$ indicate modes related to the Alice and Bob, respectively. Then, state seen by Alice and Bob experiencing Hawking radiation will be given by expanding Bob's states into Kruskal modes. Thus, new basis for the $\rho_{A B_I B_{II}}$ state is $\ket{000}, \ket{001}, \ket{010}, \ket{011}, \ket{100}, \ket{101}, \ket{110}, \ket{111}$ where $\ket{abc}=\ket{a }_A \ket{b}_{B_I}\ket{c}_{B_{II}}$. Such procedure is standard for describing similar setups in the field of RQI \cite{alsing2004a, alsing2004b, alsing2006, he2016, pan2008, ge2008}. Note, that Bob's transition to the Kruskal vacuum is analogous to the Minkowski-Rindler transition studied before in literature \cite{alsing2004a,alsing2004b,tian2013}. To further backup the consistency of such quantum states, we note that the stress-energy tensor will be regular for the Kruskal as well as the Schwarzshild coordinate system. Moreover, for the singular Boulware case, the divergence near the event horizon does not matter, since the domain of integration for the associated stress energy tensor does not includes the event horizon \cite{visser1997}. We also remark, that there is some freedom in choice of the initial Bell-type state and herein we choose the above form to allow comparison with the results presented in \cite{tian2013,he2016}.

In order to obtain physically accessible correlations of interest, we have to trace over the region $II$ which is causally disconnected from the exterior. Such trace over the region $II$ gives:
\begin{align}
    \rho_{AB_I}=\begin{pmatrix}\frac{1}{4}\alpha^2(1+\eta) & 0 &0 &\frac{1}{2}\alpha \eta\\
    0 & \frac{1}{4}\beta^2(1+\eta) &0&0  \\0&0&0 &0
    \\\frac{1}{2}\alpha \eta&0&0& \frac{1}{4}(1+\eta)
    \end{pmatrix}.
    \label{eq29}
\end{align}
For the bipartite state $\rho$, which is shared by $A$ and $B$, the MIN is defined by the following expression:
\begin{align}
   \text{MIN}(\rho)=\text{Max}_{\prod^ A}\mid\mid \rho- \prod^A (\rho)\mid\mid^2.
   \label{eq30}
\end{align}
In Eq. (\ref{eq30}), the maximum is taken over the local von Neumann measurements $\prod^A=\{\prod_l^A\}$ $(l=1,2)$ that do not disturb $\rho^A$ locally. It means that $\prod_k^A\rho^a\prod_k^A=\rho^A$ and the norm of the states is arbitrary, depending on the context. In this work, we will use the Hilbert-Schmidt norm $\mid\mid x^2\mid \mid:= \text{tr}X^\dagger X$. Note, however, that the MIN measure based on the trace norm has some potential shortcomings (see \cite{wu2014,hu2018} for more details).


\begin{figure}[ht!]
\includegraphics[width=\textwidth]{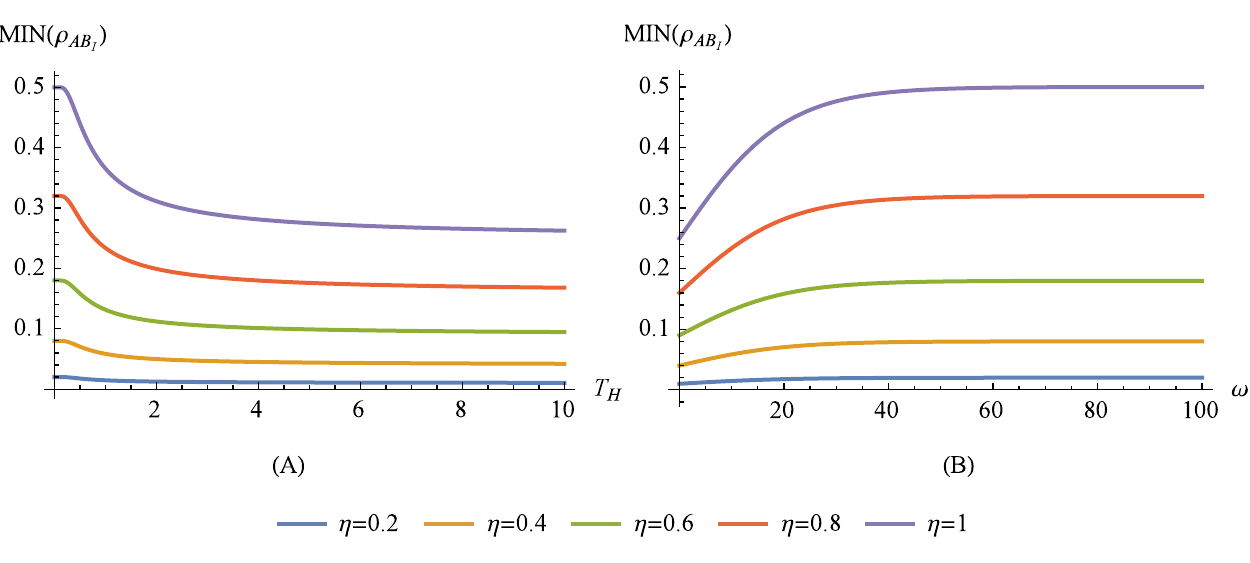}
\caption{The measurement-induced nonlocality (MIN) as a function of (A) the Hawking temperature ($T_H$) and (B) the fermionic frequency ($\omega$) for the physically accessible correlations in the case of the fermionic modes ($\rho_{AB_{I}}$). The results are presented for the selected values of the entanglement parameter ($\eta$), assuming (A) the fixed fermionic frequency ($\omega=1$) or (B) the fixed Hawking temperature ($T_H=10$).}
\label{fig2}
\end{figure}


To proceed, it is convenient to define arbitrary states composed of the two qubits in the Bloch decomposition:
\begin{align}
    \rho=\frac{1}{4}\Big( I \otimes I &+ \sum_i x_i \sigma_i\otimes I + \sum_i y_i \otimes I \sigma_i+\sum_{i,j}t_{ij} \sigma_i \otimes \sigma_j \Big),
    \label{eq31}
\end{align}
where $I$ is the $2\times 2$ identity matrix and $\sigma_i$ are the Pauli matrices ($i,j \in \{1,2,3\}$) with $x_1,x_2,x_3$ and $y_1,y_2,y_3$ being the Bloch vectors and $t_{ij}=\text{tr}[\sigma_i \otimes \sigma_j\rho^X]$ denoting the tensor of correlation. We notice that Eq. (\ref{eq29}) is of the following X-shape form:
\begin{align}
    \rho^X=\begin{pmatrix}
    \rho_{11} & 0 & 0& \rho_{14}\\
    0& \rho_{22} &\rho_{23} &0\\
    0& \rho_{32}& \rho_{33}& 0\\
    \rho_{41}& 0 & 0 & \rho_{44}\end{pmatrix},
    \label{eq32}
\end{align}
where $\rho_{ij}$ are the real parameters for the two-qubit state. We can express the parameters of state $\rho^X$ of the Bloch decomposition as below:
\begin{align}\nonumber
    &x_1=x_2=y_1=t_{12}=t_{21}=t_{13}=t_{31}=t_{23}=t_{32}=0,\\\nonumber
    &x_3=\text{tr}(\sigma^A_z \rho^X)=\rho_{11}+\rho_{22}-\rho_{33}-\rho_{44},\\ \nonumber &y_3=\text{tr}(\sigma_z^B\rho^X)=\rho_{11}-\rho_{22}+\rho_{33}-\rho_{44},\\\nonumber
    &t_{11}=\text{tr}(\sigma_x^A\sigma^B_x \rho^X)=2\rho_{14}+2\rho_{23},\\ \nonumber
     &t_{22}=\text{tr}(\sigma^A_y \sigma^B_y \rho^X)=- 2\rho_{14}+2\rho_{23},\\&
     t_{33}=\text{tr}(\sigma^A_z \sigma^B_z \rho^X)=\rho_{11}-\rho_{22}-\rho_{33}+\rho_{44}.   
     \label{eq33}
\end{align}
In accordance with the Theorem 3 presented in \cite{luo2011}, the MIN of the two-qubit states $X$ can be given by:
\begin{align}
    \text{MIN}_A (\rho^X)=
\left\{ \begin{array}{ll}
        \frac{1}{4}(t_{11}^2+t_{22}^2) & \mbox{for $x\neq 0$}\\
\frac{1}{4}(t_{11}^2+t_{22}^2+t_{33}^2-\delta_{\text{min}})  & \mbox{for  $x=0$},\end{array} \right. 
 \label{eq34}
\end{align}
where $\delta_{\text{min}}=\text{min}\{t_{11}^2,t_{22}^2,t_{33}^2\}$.
Applying this method to Eq. (\ref{eq29}) we arrive at the MIN for the physically accessible correlations:
\begin{align}
\text{MIN}(\rho_{AB_I})=\frac{\eta^2}{2(1+e^{-\omega /T_H})}.
 \label{eq35}
\end{align}
Note that the MIN will depend on the parameter $\eta$ as well as on the Hawking temperature, thus being sensitive to the parameters of a BH (mass, charge \textit{etc}.). This fact is presented in Fig. (\ref{fig2}) (A) where the behavior of the MIN as a function of $T_H$ for the specific values of $\eta$ is depicted. Analogously to the MIN studied for the Unruh effect in \cite{tian2013}, one can observe there that the MIN will not be vanishing since $\lim_{T_H \rightarrow \infty}\text{MIN}(\rho_{AB_I}= \frac{\eta}{4})$. This statement holds for the entire range of the $\eta$ parameter. It is additionally worth to remark that for $\text{MIN}(\rho_{AB_I}) \leq 0.25$ the Bell's inequality is obeyed, whereas for the $\text{MIN}(\rho_{AB_I}) >0.25$ it is violated, leading to the nonlocal quantum correlations \cite{he2016}. This is to say, the violation of the Bell's inequality as the Hawking temperature increases will be $\eta$-dependent. Hence, it will depend on the choice of the initial state. On the other hand, the case when the $T_H$ is fixed is presented in Fig. (\ref{fig2}) (B) for the selected values of the parameter $\eta$. Therein, increase of the fermionic frequency clearly leads to the increase of the $\text{MIN}(\rho_{AB_I})$, thus high values of the MIN can be sustained. This can also be observer when noting that $\lim_{\omega \rightarrow \infty}\text{MIN}(\rho_{AB_I})=\frac{\eta}{2}$.


\subsection{Bosonic modes}


The starting point of the analysis of the MIN for bosons is the following Bell state:
\begin{align}
    \ket{\phi^+}
    &=\frac{1}{\sqrt{2}}(\ket{00}+\ket{11})=\frac{1}{\sqrt{2}}\Big(\sqrt{1-t^2}\sum_{n=0}^\infty t^n \ket{0,n_I,n_{II}}+(1-t^2)\sum_{n=0}^\infty \sqrt{n+1}t^n\ket{1,n+1_I,n_{II}}\Big)
    \label{eq36}
\end{align}
In this context, to study the MIN for the bosonic state given by the $\rho_{AB_I}$, one needs to use projective measurement on the Alice's state. Such measurement can be taken over the  single-qubit system. Thus, it is possible to parametrize it by the unit vector $\vec{x}=(x_1,x_2,x_3)$ with the aid of the projectors \cite{brown2012}:
\begin{align}
    \Pi_\pm = \frac{1}{2}(I \pm \vec{x}\cdot \vec{\sigma}).
     \label{eq37}
\end{align}
Next, these projectors can be rewritten in the following form:
\begin{align}
    \Pi_{\pm}&=\frac{1}{2}\Big[(1\pm x_3)\ket{0}\bra{0}+(1\mp x_3)\ket{1}\bra{1}\pm (x_1 -ix_2)\ket{0}\bra{1} \pm (x_1+i x_2)\ket{1}\bra{0}  \Big].
\label{eq38}
\end{align}
Based on the Eq. (\ref{eq38}) one can get the post-measured final state:
\begin{align}
    \rho_{A B_I}'=\sum_{\alpha =\pm} (\Pi_\alpha \otimes I) \rho_{A B_I}(\Pi_\alpha \otimes I)=\sum_{\alpha =\pm} p_a \Pi_\alpha \otimes \rho_{B_I|\alpha},
    \label{eq39}
\end{align}
where:
\begin{align}
    \rho_{B_I|\alpha}\equiv\Tr\big((\Pi_{\alpha}\otimes I_B) \rho_{AB_I}(\Pi_{\alpha}\otimes I_B)\big)/p_{\alpha},
    \label{eq40}
\end{align}
is the post-measured state of the Bob's system conditioned on the outcome $\alpha$ with the probability $p_{\alpha}$. Also with the aid of Eq. (\ref{eq38}) one can obtain ($p_{\pm}=1/2$) \cite{brown2012}:
\begin{align}
    \rho_{B_I| \pm}\equiv \frac{1-t^2}{2} \Tilde{\rho}_I,
    \label{eq41}
\end{align}
where:
\begin{align}
    \Tilde{\rho}_{I \pm}&=(1\pm x_3)M_{00}+(1\mp x_3)M_{11}\pm(x_1-ix_2)M_{01}\pm(x_1 + ix_2)M_{10},
    \label{eq42}
\end{align}
and with the following matrices for the Bob's Hilbert space:
\begin{align}\nonumber
    M_{00}&=\frac{1}{2} (\eta+1) t^{2 n}\ket{n}\bra{n},\\ \nonumber
    M_{10}&=\eta \sqrt{n+1} \sqrt{1-t^2} t^{2 n} \ket{n+1}\bra{n}, \\ \nonumber
    M_{01}&=M_{10}^\dagger,\\
    M_{11}&=\frac{1}{2} (\eta+1) (n+1) \left(1-t^2\right) t^{2 n}\ket{n+1}\bra{n+1}.
\label{eq43}
\end{align}
After the above preparations, we can now calculate the quantity of interest:
\begin{align}
    \Tr\big((\rho_{AB_I}- \rho'_{AB_I} )^2  \big)&= \frac{(1-t^2)^2}{4}\Big[   \Tr(X_{00}^2) +2\Tr(X_{01}X_{10})+\Tr(X_{11}^2)\Big],
    \label{eq44}
\end{align}
with the $X$'s given by:
\begin{align}\nonumber
    X_{00}&\equiv M_{00}-\frac{1}{4}[(1+x_3)\Tilde{\rho}_++(1-x_3)\Tilde{\rho}_-],\\ \nonumber
    X_{11}&\equiv M_{11}-\frac{1}{4}[(1-x_3)\Tilde{\rho}_++(1+x_3)\Tilde{\rho}_-], \\ \nonumber
    X_{01}&\equiv M_{01}-\frac{1}{4}(x_1-ix_2)(\Tilde{\rho}_+-\Tilde{\rho}_-),\\
    X_{10}&\equiv M_{10}-\frac{1}{4}(x_1+ix_2)(\Tilde{\rho}_+-\Tilde{\rho}_-).
    \label{eq45}
\end{align}

We note that traces of the $X$'s will be given by the linear combinations of the matrices $M$, reducing $ \Tr\big((\rho_{AB}- \rho'_{AR} )^2\big)$ to \cite{brown2012}:
\begin{align}
     \Tr\big((\rho_{AB_I}- \rho'_{AB_I} )^2\big)&=\frac{(1-t^2)^2}{8}[(1-x_3^2)\big(\Tr(M_{00}^2)+\Tr(M_{11}^2)-2 \Tr(M_{00}M_{11})\big)+2(1+x_3^2)\Tr(M_{01}M_{10})\big)],
     \label{eq46}
\end{align}
with their traces equal to:
\begin{align}
    \Tr(M_{00}^2)=\frac{1}{4}(1+\eta)^2 \sum_{n=0}^\infty t^{4n},\;\;\;\;
    \Tr(M_{01}M_{10})=\frac{\eta^2(1-t^2)}{(t^4-1)^2},\;\;\;\;
    \Tr(M_{11}^2)= -\frac{(\eta+1)^2 \left(t^4+1\right)}{4 \left(t^2-1\right) \left(t^2+1\right)^3}.
    \label{eq47}
\end{align}
For the $x_3=1$, one can finally obtain the MIN for the bosonic case:
\begin{align}
     \Tr\big((\rho_{AB_I}- \rho'_{AB_I} )^2\big)=   \frac{\eta^2 \left(1-t^2\right)^2 \left(t^2-1\right)^3 \left(t^4+1\right)}{8
   \left(t^4-1\right)^3}.
   \label{eq48}
\end{align}
We remark that setting $x_3=0$ in Eq. (\ref{eq39}) leads to the quantum discord \cite{brown2012,tian2013}.


\begin{figure}[ht!]
\includegraphics[width=\textwidth]{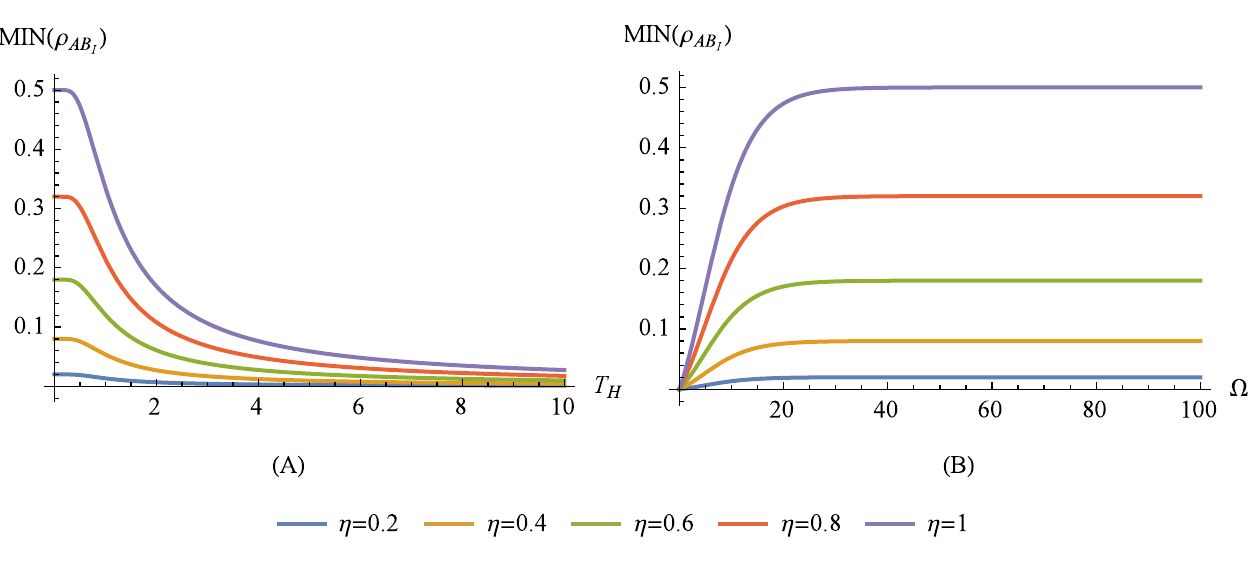}
\caption{The measurement-induced nonlocality (MIN) as a function of (A) the Hawking temperature ($T_H$) and (B) the bosonic frequency ($\Omega$) for the physically accessible correlations in the case of the bosonic modes ($\rho_{AB_{I}}$). The results are presented for the selected values of the entanglement parameter ($\eta$), assuming (A) the fixed fermionic frequency ($\Omega=1$) or (B) the fixed Hawking temperature ($T_H=10$).}
\label{fig3}
\end{figure}


The behavior of the bosonic MIN is presented in Fig. (\ref{fig3}) (A) for the Hawking temperature in the range $T_H \in (0,10)$ and with fixed $\eta\in(0,1)$. In accordance to the results presented in \cite{tian2013,he2016}, the $\text{MIN}(\rho_{AB_I})$ vanishes rapidly and $\lim_{T_H\rightarrow \infty}\text{MIN}(\rho_{AB_I})= 0$. However, by increasing the frequency of the bosonic modes it is again possible to maintain relatively high values of the MIN. This scenario is presented in Fig. (\ref{fig3}) (B) where for $T_H=10$, the frequency $\Omega$ takes values in the range $\Omega \in (0,100)$. Interestingly, $\lim_{\Omega \rightarrow \infty}\text{MIN}(\rho_{AB_I})=\frac{\eta}{2}$ and coincides with the result previously obtained for the fermionic modes with frequency $\omega$. Thus, for a high frequencies of the modes, the behavior of the bosonic and fermionic modes for the finite $T_H$ will be similar.


\subsection{Mixed boson-fermion modes}


In order to describe the mixed boson-fermion case, we slightly change setup used in the previous sections. In particular, Alice is still equipped with the fermionic mode detector, while Bob can detect bosons. However, Alice, just like Bob after the free fall, will now hover near the horizon of a BH. Both of them will detect Hawking temperature in their corresponding detectors. An analogous situation but with two accelerated observers and in the different context can be found in \cite{richter2015} or for tripartite system with two noninertial observers in \cite{wang2010}. In the case of both parties experiencing Hawking radiation, the fermionic modes of Alice in the black hole coordinates will be given via Eq. (\ref{eq19}), while the bosonic modes of Bob are written according to Eq. (\ref{eq26}):
\begin{align}\nonumber
    \ket{\phi^+}&=\frac{1}{\sqrt{2}}(\sqrt{1-t^2}\sum_{n=0}^\infty t^n \ket{n_I, n_{II},1_{I},0_{II}})\\ &+(1-t^2)\sum_{n=0}^\infty \sqrt{n+1}t^n\ket{n+1_I,n_{II}}(\alpha \ket{0_I0_{II}}+\beta \ket{1_I1_{II}}).
    \label{eq49}
\end{align}
In correspondence to the previous cases, state $\rho_{A_I A_{II}B_I B_{II}}$ have to be traced over region $II$, which is inaccessible for both Alice and Bob. The obtained density matrix $\rho_{A_IB_I}$ will be treated in a manner similar to the Eqs. (\ref{eq37}) - (\ref{eq48}). We would like to note that for the singlet Bell state, $\ket{\psi^-}=\frac{1}{\sqrt{2}}(\ket{01}-\ket{10})$ it is not possible to calculate MIN based on the trace norm \cite{luo2011}. Terms required to obtain the MIN given by Eq. (\ref{eq39}) are vanishing for the $x_3=1$. In what follows, the coefficient of the decomposition will have the following shape:
\begin{align}\nonumber
    M_{00}&=\sum_{n=0}^\infty \alpha^2(1+\eta) \frac{1}{2}t^{2n}\ket{n}\bra{n},\\ \nonumber
    M_{01}&=\sum_{n=0}^\infty \sqrt{1+n} \sqrt{1-t^2}\alpha\eta t^{2n}\ket{n}\bra{n+1}, \\ \nonumber
    M_{10}&=\sum_{n=0}^\infty \sqrt{1+n} \sqrt{1-t^2}\alpha \eta t^{2n}\ket{1+n}\bra{n}, \\ M_{11}&=\sum_{n=0}^\infty\frac{1}{2}(1+\eta)(1+n)(1-t^2)t^{2n}\ket{1+n}\bra{1+n}+\frac{1}{2}(1+n)t^{2n}(1+\eta)\beta^2\ket{n}\bra{n},
    \label{eq50}
\end{align}
where $\alpha$ and $t$ are specified by Eqs. (\ref{eq18}, \ref{eq24}, \ref{eq25}). Hence:
\begin{align}\nonumber
    \Tr(M_{00}^2)&=\frac{\alpha^4}{4(1-t^4)}(1+\eta),\\ \nonumber
    \Tr(M_{01}M_{10})&=\frac{\alpha ^2 \eta^2 \left(1-t^2\right)}{\left(t^4-1\right)^2},\\ \nonumber \Tr(M_{00}M_{11})&=\frac{\alpha ^2 (\eta+1)^2 \beta ^2}{4 \left(t^4-1\right)^2},\\
    \Tr(M_{11}^2)&=-\frac{(\eta+1)^2 \beta ^4 \left(t^4+1\right)}{4 \left(t^4-1\right)^3}-\frac{(\eta+1)^2 \left(t^4+1\right)}{4 \left(t^2-1\right) \left(t^2+1\right)^3}.
    \label{eq51}
\end{align}
Assuming $x_3=1$, one gets:
\begin{align}
   \Tr\big((\rho_{A_IB_I}- \rho'_{A_I B_{I}} )^2\big)= \frac{1}{2}
   \left(1-t^2\right)^2\Big(        -\frac{\alpha ^2 \eta^2
   \left(t^2-1\right)
   }{
   \left(t^2+1\right)^2}\Big).
   \label{eq52}
\end{align}
%


\begin{figure}[ht!]
\includegraphics[width=\textwidth]{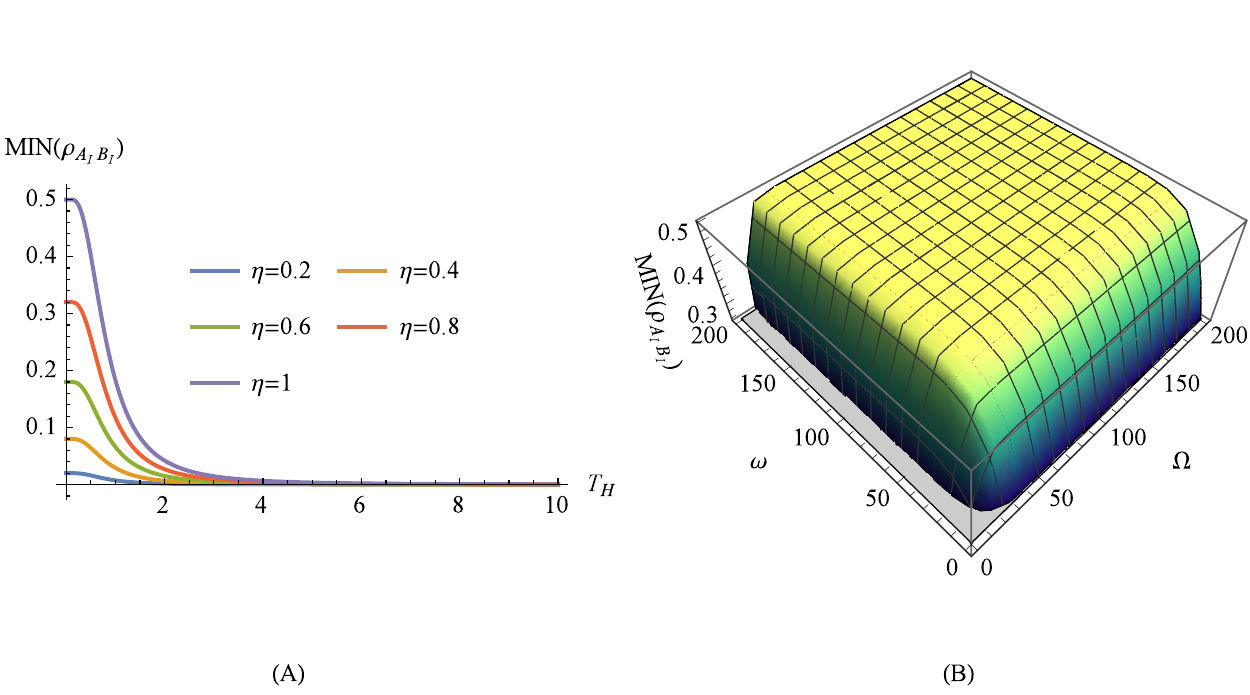}
\caption{The measurement-induced nonlocality (MIN) as a function of (A) the Hawking temperature and (B) simultaneously the fermionic ($\omega$) and bosonic ($\Omega$) frequencies for the physically accessible correlations in the case of the mixed modes ($\rho_{A_{I}B_{I}}$). The results are presented for (A) the selected values of the entanglement parameter ($\eta$) and (B) for the fixed entanglement parameter ($\eta=1$) and the fixed Hawking temperature ($T_H=10$).}
\label{fig4}
\end{figure}


The MIN behavior for the boson-fermion case as a function of the Hawking temperature for different values of the parameter $\eta$ is presented in Fig. (\ref{fig4}) (A). Interestingly, for the small Hawking temperature, the MIN initially behaves similar to the fermionic case. Then, the MIN will quickly decrease, leading to the $\lim_{T_H \rightarrow \infty}=0$. As the $T_H$ increases, the bosonic contribution will dominate. An interesting situation is depicted in the Fig. (\ref{fig4}) (B), where for the fixed $T_H=10$ and $\eta=1$ we plot the behavior of the $\text{MIN}(\rho_{AB_I})$ as a function of the bosonic ($\Omega$) and fermionic ($\omega$) frequencies, respectively. Just like in the fermionic case, by increasing $\omega$ it is possible to achieve rather high values of the $\text{MIN}(\rho_{A_IB_I})$.


\subsection{Physically inaccessible correlations}


The conducted analysis can be further supplemented by the discussion of the physically inaccessible correlations. Specifically, to describe the possibility of the destruction or the transfer of correlations, in the following part we trace over the region $B_I$. In this way, the MIN for the physically inaccessible correlations can be obtained \cite{he2016}.

For the fermion-fermion state, we obtain the following density matrix:
\begin{align}
    \rho_{AB_{II}}=\begin{pmatrix}
\frac{1}{4} \alpha^2(1+\eta) & 0 & 0 & 0 \\
0 & \frac{1}{4}\beta^2(1+\eta) & \frac{1}{2}\eta\beta & 0 \\
0 & \frac{1}{2}\eta\beta & \frac{1}{4}(\eta+1) & 0 \\
0 & 0 & 0 & 0 
\end{pmatrix}.
\label{eq53}
\end{align}
Next, the steps analogous to the ones conducted for the physically accessible correlations, but with the partial trace over region $B_I$ instead of region $B_{II}$, lead to the $\text{MIN}(\rho_{AB_{II}})$:
\begin{align}
  \text{MIN}(\rho_{AB_{II}})=  \frac{\eta^2}{2 \left(e^{\frac{\omega }{T_H}}+1\right)}.
  \label{eq54}
\end{align}
On the other hand, for the boson-fermion situation we have the following matrices $M$:
\begin{align}
    M_{00}&=\sum_{n=0}^\infty\Big[\frac{1}{2}t^{2n}\alpha^2 \ket{n}\bra{n}+ \frac{1}{2}(1+n)t^{2n}(1-t^2)(1+\eta)\ket{1+n}\bra{1+n}\Big],\\ \nonumber
    M_{11}&=\sum_{n=0}^\infty\frac{1}{2}(1+n)t^{2n}(1+\eta)\beta \ket{n}\bra{n},\\ \nonumber
    M_{01}&=\sum_{n=0}^\infty(1+n)t^{2n}\sqrt{1-t^2}\eta\beta \ket{n+1}\bra{n},\;\;\; M_{10}=M_{01}^\dagger.
    \label{eq55}
\end{align}
Straightforwardly repeating the procedure presented in Eqs. (\ref{eq37})-(\ref{eq48}), with the trace over $A_{II}$ and $B_I$, one can arrive at the $\text{MIN}(\rho_{A_IB_{II}})$ of the mixed state:
\begin{align}
   \text{MIN}(\rho_{A_IB_{II}})= \frac{ \left(1-t^2\right)^2 \left(\eta^2 \beta ^2 \left(1-t^2\right) \left(-t^4-1\right)\right)}{2 \left(t^4-1\right)^3}.
\end{align}
%


\begin{figure}[ht!]
\includegraphics[width=\textwidth]{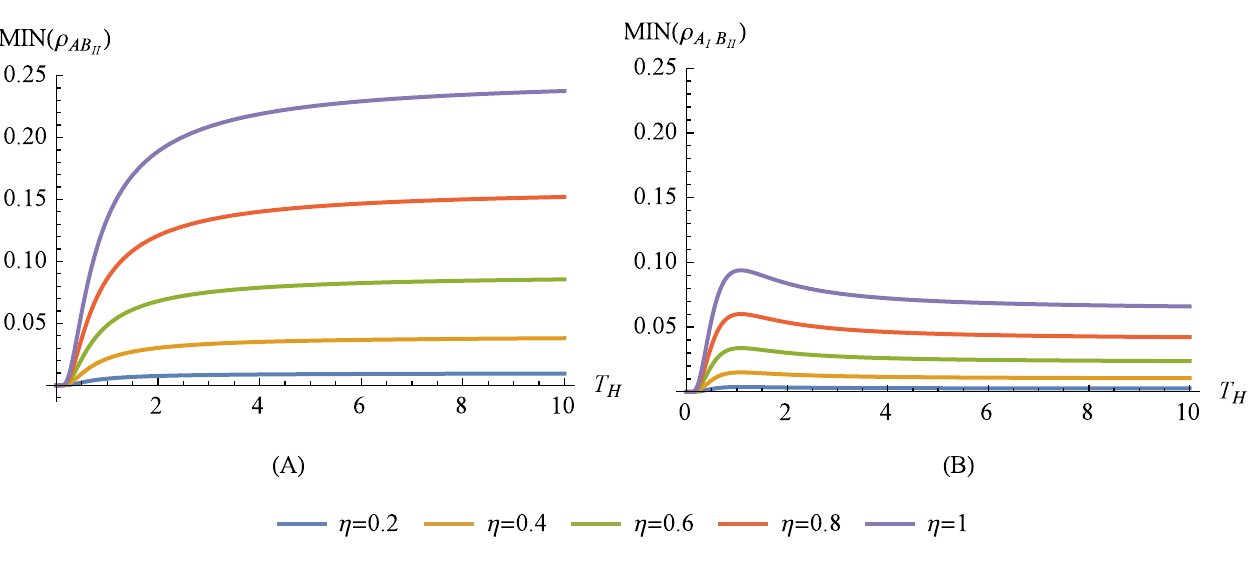}
\caption{The measurement-induced nonlocality (MIN) as a function of the Hawking temperature ($T_{H}$) for the physically inaccessible correlations in the case of (A) the fermionic and (B) the mixed modes. The results are presented for the selected values of the entanglement parameter ($\eta$) and for the fixed fermonic ($\omega=1$) and bosonic frequencies ($\Omega=1$).}
\label{fig5}
\end{figure}


By inspecting Fig. (\ref{fig5}) (A), one can observe that the physically inaccessible correlations increase under the effect of the Hawking radiation for the fermionic case, contrary to the previously analyzed region $AB_I$ with the $\lim_{T_H\rightarrow\infty}\text{MIN}(\rho_{A_IB_{II}})=\eta^2/4$. On the contrary, in the mixed fermion-boson case, the result will be finite for the infinite Hawking temperature, $\lim_{T_H\rightarrow\infty}\text{MIN}(\rho_{A_IB_{II}})=\eta/16$, as depicted in Fig. (\ref{fig5}) (B). Moreover, increase of the frequencies $\omega$ for the Dirac modes leads to the destruction of the correlations for the physically inaccessible correlations. However, this is not the case for the mixed fermion-boson correlations for the region $A_I B_{II}$, where $\lim _{\omega,\Omega\rightarrow \infty}\text{MIN}(\rho_{A_IB_{II}})=\eta^2/2(1 + e^{(1/T_H)})$. The results presented here confirm that nonlocality is highly dependent on the choice of the modes.

\section{Concluding remarks and future perspectives}


In this work, we have presented extended and unified discussion of nonlocality for the fermonic and bosonic fields under relativistic effects. Similarly to the discussions presented in \cite{tian2013,he2016}, the nonlocality was investigated via the quantum measure known as the measurement-induced nonlocality (MIN) \cite{luo2011}. However, by using the generic black hole (BH) metric with the corresponding Hawking temperature ($T_{H}$), we were able to study general properties of the MIN in the {\it model-independent} way. Thereby, it was possible to gain a broad insight into the behavior of different excitations on the same footing and to establish comprehensive context for future investigations.

In details, it was found that the MIN in the fermionic case takes finite value (equal to $\eta/2$, where $\eta$ is the entanglement parameter) when $T_{H}$ approaches infinity. On the other hand, the MIN for the bosonic and mixed modes equals to zero at the same infinite-temperature limit, indicating that the corresponding quantum correlations vanish. This is to say, fermions appears to be {\it more robust} toward the Hawking radiation than the bosons, in analogy to the resistance against acceleration as discussed in \cite{richter2015}. Strictly speaking, the reason behind such behavior is that the bosonic and mixed correlations will vanish from the unboundness of the bosonic partition function that composes the MIN. Nonetheless, in addition to the above, we have shown that for the higher mode frequencies it is still possible to sustain high values of the MIN for the finite Hawking temperature and physically accessible correlations. In particular, the MIN for the bosonic and fermionic modes were observed to coincide to the finite $\eta/2$ value at the infinite-frequency limit, being independent of the Hawking temperature. Thus, the modes characterized by the relatively high vibrational frequencies (energies) appears to be favored in terms of preserving quantum correlations. We find that this behavior is analogous in the case of the mixed modes. For convenience, our results are summarized in the Table \ref{table1}. Note that these findings partially confirm previous specific studies (see \cite{tian2013,he2016}) and for the first time provide unified picture of the interplay between different mode types in the context of the MIN.

In fact the discussed effects are the subject for further inspection based on the mixed mode case, hitherto not considered in the literature in terms of the MIN. In general, the behavior of MIN for the mixed modes seems to be qualitatively closer to the bosonic case, as described above. Hence, it can be stated that the bosonic part of the mixed mode takes dominant role in shaping the nonlocality behavior near a black hole. This suggests the negative effect of the Bose-Einstein statistics on the quantum correlations related to the nonlocality. Such observation is reinforced by the results obtained here for the physically inaccessible correlations. Specifically, the fermionic MIN tends to be partially transferred from the region $I$ to region $II$ and the magnitude of this transfer increases along with the temperature. However, for the mixed modes the amount of the MIN transferred from the region $I$ to $II$ is significantly lower. It can be then concluded that when the MIN for the physically accessible modes vanishes, the remaining correlations are stored in the inaccessible correlations. In what follows, the impact of the bosonic modes on correlations appears to be much stronger than previously anticipated \cite{richter2015}.

Moreover, we note that the MIN may change rather arbitrarily through uncorrelated action of the unmeasured part. Therefore, we would like to remark that there are other forms of the MIN as well as the alternative measures of nonlocality \cite{hu2012,xi2012,li2016,muthagenesan2017}. Potential study of such quantifiers in the relativistic setups may shed a new light on the details and the conceptual differences between existing forms of correlation. In this manner, our study can be considered as a reference point for future comparisons with various quantum measures within the general theoretical framework adopted here. For example, since nonlocality is a broad concept and there may be {\it quantum nonlocality without entanglement} \cite{walgate2002} and {\it quantum nonlocality without quantumness} \cite{bennet1996}, the quantum measures of nonlocal phenomena are expected to capture different correlations than those based on the entanglement and quantumness \cite{luo2011}. However, other forms of measuring nonlocality may be also distinct from each other due to the asymmetry of the \textit{gedanken} setup used here, in analogy to the discord-like form of correlations affected by the Unruh effect \cite{celeri2010}. In other words, no universal measure of nonlocality may be available for a systems affected by the black hole geometry. Note that such predicted direction of research is certainly intriguing not only in terms of the four-dimensional static black holes but also when considering their higher-dimensional or rotating counterparts, which were already extensively studied in other contexts \cite{fuentes2005,ge2008,pan2008,he2016}.

Lastly, since our work employs single-mode approximation, the approach that is still raising some concerns among researchers, it would be worthwhile to extend the analysis of nonlocality measures in relativistic context beyond this approximation \cite{bruschi2010,hu2018}. By doing so, one may establish not only a more refined theoretical model but also transfer the entire analysis from the \textit{thought} experiments to a somewhat realistic settings. This is also related to the assumption when detectors are considered localized and \textit{see} specific modes of the field. In order to work with this kind of observers, the specific form of coupling between field and the detector ought to be considered.  Some works already are pointed in that direction, with reference to the cavities or the Unruh-DeWitt detectors \cite{celeri2010,Lee2014}. Another possible approach may be even more fundamental and related to solving the Dirac or (and) Klein-Gordon equations with a smaller number of assumptions and approximations \cite{crucean2023}. Moreover, introduction of localized projective measurements may properly describe localized modes of quantum fields in curved spacetimes \cite{Dragan2013,dragan2013b}. As it was shown, detector of this kind may be used for projective measure of the single mode of quantum field, resolving some issues with the \textit{standard} single-mode approach. As a summary, work presented here also attempts to gain insight in the setup-building procedure, which the vast literature on RQI still lacks \cite{ge2008,pan2008,wang2010,tian2013,he2016}. The perspectives listed above should be properly addressed, which will be the next task of the authors in the ongoing research.

\begin{table}[]
\begin{tabular}{|c|c|c|c|}
 \hline
 &   &   &  \\
 & fermionic  & bosonic  & mixed \\
 & modes  & modes  & modes \\
 &   &   &  \\\hline
 &   &   &  \\
 $\lim _{T_H\rightarrow \infty}\text{MIN}(\rho_{A B_I})$ & $\eta/4$ & $0$ & $-$ \\
  $\lim _{T_H\rightarrow \infty}\text{MIN}(\rho_{A_I B_I})$& $-$ &$-$  & $0$ \\
  $\lim _{\omega,\Omega\rightarrow \infty}\text{MIN}(\rho_{A B_I})$& $\eta/2$ & $\eta/2$ & $-$ \\
  $\lim _{\omega,\Omega\rightarrow \infty}\text{MIN}(\rho_{A_I B_I})$& $-$ & $-$ & $\eta/2$ \\
$\lim _{T_H\rightarrow \infty}\text{MIN}(\rho_{A B_{II}})$& $\eta^2/4$ & $-$  &$-$  \\
 $\lim _{\omega,\Omega\rightarrow \infty}\text{MIN}(\rho_{A B_{II}})$& $0$  &$-$  &$-$ \\ $\lim _{T_H\rightarrow \infty}\text{MIN}(\rho_{A_I B_{II}})$& $-$ & $-$  &$\eta/16$  \\
 $\lim _{\omega,\Omega\rightarrow \infty}\text{MIN}(\rho_{A_I B_{II}})$& $-$  &$-$  &$\eta^2/2(1+e^{1/T_H})$\\
 &   &   &  \\
  \hline
\end{tabular}
\caption{The limits of the measurement-induced nonlocality (MIN) for the accessible and inaccessible correlations in the case of the fermionic, bosonic and mixed modes. In the table, $T_{H}$ stands for the Hawking temperature, $\omega(\Omega)$ is the fermionic (bosonic) frequency and $\eta$ denotes the entanglement parameter. The regions are described in details in the main text.}
\label{table1}
\end{table}


\section*{Acknowledgements}


One of us S.K. would like to acknowledge the financial support in part by the National Science Foundation under award number 1955907.


\bibliographystyle{apsrev}
\bibliography{bibliography}

\end{document}